\documentclass{SciPost}

\hypersetup{
    colorlinks,
    linkcolor={red!50!black},
    citecolor={blue!50!black},
    urlcolor={blue!80!black}
}

\usepackage[bitstream-charter]{mathdesign}
\usepackage{graphicx}
\usepackage{dcolumn}
\usepackage{bm}
\usepackage{xcolor}
\usepackage{subcaption}
\usepackage{physics}
\usepackage{ragged2e}
\usepackage{mwe}
\usepackage{tcolorbox}
\usepackage{tkz-euclide}
\usepackage{wasysym}
\usepackage{lmodern}
\usepackage[T1]{fontenc}
\usepackage{hyperref}
\usepackage{blindtext, rotating}
\DeclareCaptionJustification{justified}{\justifying}
\renewcommand{\vec}[1]{{\boldsymbol #1}}

\hypersetup{
	colorlinks=true,
	urlcolor=blue
}
\DeclareSymbolFont{usualmathcal}{OMS}{cmsy}{m}{n}
\DeclareSymbolFontAlphabet{\mathcal}{usualmathcal}

\fancypagestyle{SPstyle}{
\fancyhf{}
\lhead{\colorbox{scipostblue}{\bf \color{white} ~SciPost Physics }}
\rhead{{\bf \color{scipostdeepblue} ~Submission }}

\fancyfoot[C]{\textbf{\thepage}}
}

\begin{document}

\pagestyle{SPstyle}

\begin{center}{\Large \textbf{\color{scipostdeepblue}{
Diffusion and relaxation of topological excitations in layered spin liquids\\
}}}\end{center}

\begin{center}\textbf{
Aprem P. Joy\textsuperscript{$\star$},
Roman Lange and
Achim Rosch
}\end{center}

\begin{center}
Institute for Theoretical Physics, University of Cologne, Cologne, Germany
\\
[\baselineskip]
$\star$ \href{mailto:email1}{\small aprempjoy@gmail.com}\,,\quad
\end{center}

\section*{\color{scipostdeepblue}{Abstract}}
\textbf{\boldmath{%
Relaxation processes in topological phases such as quantum spin liquids are controlled by the dynamics and interaction of fractionalized excitations. In layered materials hosting two-dimensional topological phases, elementary quasiparticles can diffuse freely within the layer, whereas only pairs (or more) can hop between layers - a fundamental consequence of topological order. Using exact solutions of emergent nonlinear diffusion equations and particle-based stochastic simulations, we explore how pump-probe experiments can provide unique signatures of the presence of $2d$ topological excitations in a $3d$ material. Here we show that the characteristic time scale of such experiments is inversely proportional to the initial excitation density, set by the pump intensity.
A uniform excitation density created on the surface of a sample spreads subdiffusively into the bulk with a mean depth $\bar z$ scaling as $\sim t^{1/3}$ when annihilation processes are absent. The propagation becomes logarithmic, $\bar z  \sim \log t$, when pair-annihilation is allowed. Furthermore, pair-diffusion between layers leads to a new decay law for the total density, $n(t) \sim (\log^2 t)/t$ - slower than in a purely $2d$ system. 
We discuss possible experimental implications for pump-probe experiments in finite-size system.
}}

\vspace{\baselineskip}

\noindent\textcolor{white!90!black}{%
\fbox{\parbox{0.975\linewidth}{%
\textcolor{white!40!black}{\begin{tabular}{lr}%
  \begin{minipage}{0.6\textwidth}%
    {\small Copyright attribution to authors. \newline
    This work is a submission to SciPost Physics. \newline
    License information to appear upon publication. \newline
    Publication information to appear upon publication.}
  \end{minipage} & \begin{minipage}{0.4\textwidth}
    {\small Received Date \newline Accepted Date \newline Published Date}%
  \end{minipage}
\end{tabular}}
}}
}


\vspace{10pt}
\noindent\rule{\textwidth}{1pt}
\tableofcontents
\noindent\rule{\textwidth}{1pt}
\vspace{10pt}


\section{Introduction}

Topological phases of matter—such as fractional quantum Hall states and quantum spin liquids—host emergent quasiparticles that carry fractional quantum numbers and obey non-trivial exchange statistics. Their spatially non-local character makes them highly appealing for quantum information processing, yet this same feature renders them exceptionally difficult to detect experimentally. Conventional probes, such as neutron spectroscopy, couple to spatially local operators and therefore typically excite a broad continuum of states \cite{fieldinduced,proximate,Han2012,sandilands2015scattering,joyRaman,knolle}. As a result, direct signatures of topological order are often inaccessible in measurements based on such local observables. Identifying experimental strategies that provide unambiguous evidence of topological order and fractionalization thus remains a central challenge in modern condensed matter physics.

A promising direction is to move beyond linear response and exploit genuinely out-of-equilibrium experiments—such as pump–probe protocols—that track, with high temporal and spatial resolution, the equilibration dynamics following a sudden excitation of the system\cite{gedikpumpprobe,quenchspinice}. Recent works have shown that coherent nonlinear spectroscopy, using sequences of ultrafast laser pulses, can resolve continua associated with fractionalized excitations in quantum spin chains \cite{armitagenonlinear,hickeynonlinear,knollenonlinear}. Even more remarkably, such nonlinear probes may reveal signatures of non-trivial (anyonic) statistics in two-dimensional spin liquids \cite{mcginley,trivedi,pollmannnonlinear}. Rapid progress in ultrafast light-matter experiments have enabled pump-probe studies of various solid state systems with some recent results uncovering anomalous relaxation dynamics of topological defects in diverse material systems \cite{topdefectsprx25,mitrano2019ultrafast,rucl3ultrafast}.

\begin{figure}[t]
	\centering
	\includegraphics[width=0.4\linewidth]{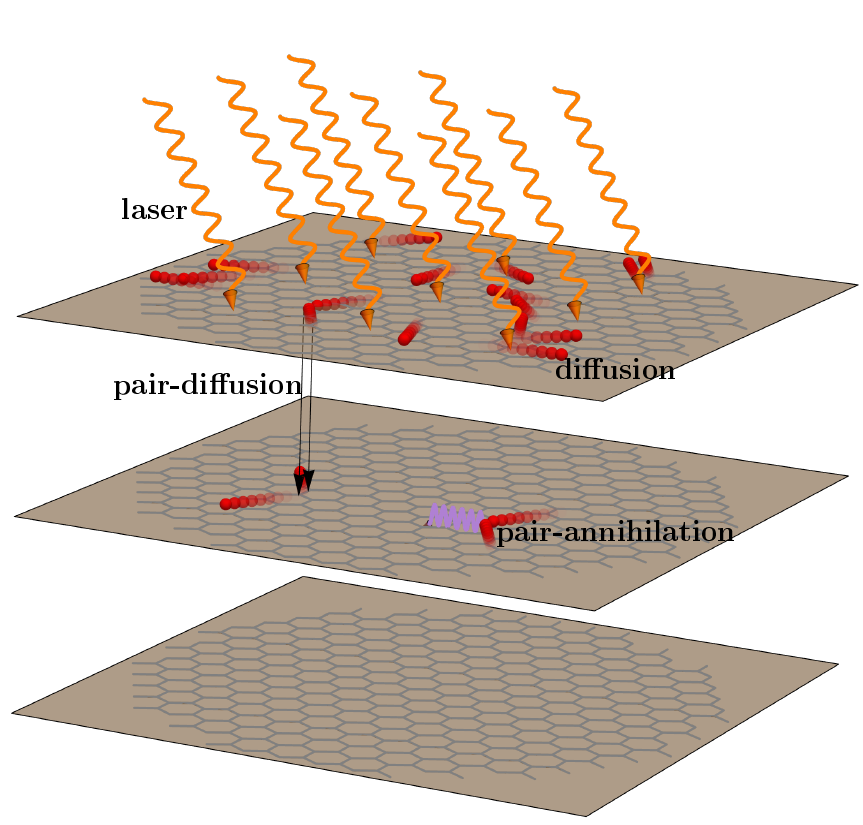}
	\caption{A layered material, e.g., $\alpha$-\text{RuCl}$_3$, hosting a $2d$ topological phase is uniformly excited from the top by a laser pulse of intensity $I_P$. Single excitations are topological (red spheres), e.g. visons in a Kitaev spin liquid, and can diffuse freely within the $2d$ layers, whereas inter-layer motion requires a pair of excitations due to the constraints imposed by topological order and the emergent gauge structures. Excitations are eliminated only via pair-annihilation into vacuum (with a rate $\lambda$), emitting low-energy trivial excitations such as phonons (blue wavy lines). This leads to a subdiffusive  and logarithmic spreading of the excitations into the bulk for $\lambda=0$  and $\lambda\ne0$ respectively. All characteristic time scales, measured via a probe laser for example (not shown), scales inversely with the pump intensity, $\tau \propto 1/I_P$.}
	\label{fig:illustration}
\end{figure}

In this work, we propose and explore a class of quench experiments--specifically suited for multilayered three-dimensional ($3d$) systems--that directly probe a robust consequence of topological order: the emergent dimensionality of quasiparticles\cite{floatingtop}. See Fig.~\ref{fig:illustration} for a sketch. In a $3d$ crystal described by weakly coupled $2d$ topological phases, the elementary excitations are confined to move within the two-dimensional planes while only a topologically trivial composite (a pair or more) can move in the third direction (through the bulk). As a direct consequence, they undergo anomalous diffusion into the bulk of a crystal which may be detected in a pump-probe experiment. These behaviors are in stark contrast to both ordinary diffusion and relaxation of topologically trivial quasiparticles such as magnons or phonons, making them observable hallmarks of topological order and emergent gauge theories. The emergent dimensionality of excitations can also strongly affect the inter-layer transport coefficients of charge and heat, as pointed out by Refs.~\cite{interlayerthermal,floatingtop}.

Our protocol is relevant to a broad class of layered spin-liquid candidates \cite{savary2016quantum,spinliquidreviewZhou,herbertsmithitereview,trebst2017kitaev,jackeli}. An instructive example is realized by stacks of weakly coupled Kitaev spin liquids \cite{Kitaev06}. Such models are directly relevant to van der Waal magnets such as $\alpha-\text{RuCl}_3$ where experimental results have suggested the existence of a spin liquid phase with emergent $Z_2$ gauge field and Majorana fermions \cite{kasahara1,kasahara2,trebst2017kitaev,gammaRau,jansenninterlayer,balzinterlayer}. While a single layer of the Kitaev model is integrable, perturbations arising, e.g., from Heisenberg or $\Gamma$ terms break integrability and induce an effective dynamics of its topological vison excitations \cite{prxvison,inti,Batista}. For a multilayered system to host the spin liquid phase, one needs that the layers are weakly coupled.  At the same time, the perturbations within the layer are assumed to be sufficiently weak to not destroy the topological order. In a previous publication, some of us have explored the emergent kinematically constrained motion of the visons in simplified models of layered Kitaev spin liquids \cite{joy2024gauge}.

Also pertinent to our discussion are the recently discovered fracton models, where mobility constraints of excitations play a central role. These are exotic $3d$ topological phases with subsystem symmetries or dipole conservation laws that hosts excitations which are either  completely immobile when isolated, or can move only along a subdimensional manifold as bound pairs \cite{fractonReview2019,pretkoreview,fractonelasticity,Fufracton}. While experimental realizations of fractons remain elusive, theoretical studies have uncovered anomalous hydrodynamics and slow equilibration in these models, directly arising from the mobility constraints \cite{fractonhydro,glorioso2022breakdown,glassyFracton,anomalousDiff,gliozzi2025domain,Castelnovo01012012}. From this perspective, a stack of weakly coupled $2d$ topological layers provides a realistic platform that captures several essential features of fracton-like constrained dynamics, thereby broadening the scope of our proposal.

The out of equilibrium dynamics of topological defects and domain walls also play a central role in the kinetics of symmetry breaking phase transitions in classical physics  \cite{quenchxy,brayquenchxy}. 
For example, the growth of ordered domains can be well understood in terms of the gradual elimination of topological defects characteristic of the broken-symmetry phase (e.g. lattice dislocations in the crystallization of a solid), resulting in universal power-laws when quenched though a phase transition, e.g. described by the famous Kibble-Zurek mechanism \cite{kibble,zurek1985cosmological,kibblezureksondhi}. Remarkably, in several cases, such topological defects also exhibit kinematically constrained motion \cite{HirthLothe1982,Cvetkovic11072006,doshi2021vortices}.

Given the challenge of identifying topological phases—especially quantum spin liquids—in real materials, detecting the emergent dimensionality of excitations would provide a strong experimental signature across a wide class of systems. The paper is organized as follows. In Sec.\ref{sec:model}, we introduce an effective particle-based model of diffusive topological excitations in a layered lattice model along with a coarse-grained noisy diffusion equation in the continuum limit. In Sec.\ref{sec:results}, we first present the central result of our analysis demonstrating the characteristic scaling predictions for quench experiments, and subsequently discuss in detail exact solutions for the noiseless diffusion equation in the infinite layer limit. In Sec.\ref{sec:noise}, we examine the corrections to our results arising from noise, and present our predictions for a finite sized sample in Sec.\ref{sec:finiteslab}. Finally, we discuss some important experimental considerations for our protocol. (See App.\ref{app:nontopo} for a more detailed discussion on experimental feasibility.)


\section{Model}
\label{sec:model}
We consider a model of topological phases in a layered three-dimensional crystal. Our main example is a stack of $2d$ $Z_2$ spin liquids. Very similar situations can arise for stacks of  fractional quantum Hall systems \cite{levinfisherfqh,balentsfisherfqh}. However, we will avoid extra complications arising from charge conservation and long-ranged Coulomb interaction in our discussion.

The excitations of our system are topological quasiparticles, e.g., anyons, visons, or spinons, which we assume to be gapped and mobile within each plane. 
We consider a situation where the quasiparticles scatter from some other degrees of freedom such as phonons or impurities. The precise form of the scattering mechanism is irrelevant. For our purpose, we only need that this leads to an effective diffusive motion of single quasiparticles on length scales large compared to their (inelastic or elastic) mean free path $\ell$.
Importantly, topological order implies that within each layer the quasiparticles can only be destroyed or created in pairs.

In the limit where the distance between quasiparticles is  larger than their mean free path $\ell$, we can treat them effectively as classical diffusive particles. We model them as a set of random-walkers (with a hard-core constraint) labeled by $A_{\vec r,l} $  where $\vec r$ the coordinate in the $2d$ plane and $l$ the layer index.  Intra-layer motion is described by Brownian type diffusion with a diffusion rate $\Gamma_\|$ for nearest neighbor hops. As a consequence of topological order, inter-layer hopping can only occur via close-by pairs. Thus, we consider only three processes: in-plane diffusion, pair-hopping, and pair annihilation, occuring on a cubic lattice schematically represented below.
\begin{align}
	A_{\vec r,l} &\overset{\Gamma_\|}{\longrightarrow} A_{\vec r+\vec \delta,l} & \text{in-plane diffusion}\nonumber\\
	A_{\vec r,l}+A_{\vec r+\vec \delta,l}&\overset{\Gamma_\|\Gamma_\perp}{\longrightarrow} A_{\vec r,l\pm1}+A_{\vec r+\vec \delta,l\pm1} & \text{pair hopping}\nonumber\\
	A_{\vec r,l}+A_{\vec r+\vec \delta,l}&\overset{\Gamma_\| \Gamma_\lambda}{\longrightarrow} 0 &  \text{pair annihilation}
	\label{eq:model}
\end{align}
where $\vec \delta$ are  nearest-neighbor vectors within a plane and $\Gamma_\perp$ and $\Gamma_\lambda$ are the rates for pair hopping and annihiliation, respectively. More precisely, pair hopping and pair annihilation are implemented in the following way: when two particles hop onto the same side, the pair is annihilated with probability $\Gamma_\lambda$, and with a probability $\Gamma_\perp$ it moves either one layer up or down, while in all other cases particles go back to their previous configuration (to implement a hard-core constraint).
We do not consider pair creation processes as we assume that the effective temperature of the system is small compared to the quasiparticle gap. For our simulations we have use two different initial states, one where the particles are randomly placed on the top layer and one where pairs of excitations occupying neighboring sites are  placed randomly. The latter initial conditions takes into account that the quasi particle can only be created in pairs. With the exception of the behavior at very short times, we find that both initial conditions give identical results, see App.~\ref{app:initial}. For all figures in the main text we use randomly placed particles as initial condition.

The simple model discussed above focuses on the effective diffusion of topological excitations.
For visons in a Kitaev model coupled to a thermal bath, Yang and Chern \cite{yang2025thermalquenchdynamicsvisons}
studied, using a kinetic Monte Carlo simulation, the effect of long-range forces mediated by Majorana fermions on vison-vison annihilation. For broad parameter regimes, they obtain (for single layers) results consistent with the simple diffusion model discussed above but they also identified regimes where, e.g., long-ranged attractive forces accelerate vison annihilation.

Note that, our model considers only one type of topological quasiparticles, but it can easily be generalized to multiple species. A subtle question concerns the dynamics of non-abelian excitations with extra internal degrees of freedom (e.g., Ising anyons in a chiral Kitaev liquid). In this case, the (diffusive) real-space dynamics leads to braiding, and thus a complex quantum dynamics of the internal degrees of freedom. At the same time, annihilation and pair-hopping is governed by fusion outcomes. For example, consider Ising anyons in a system with a large gap to fermionic excitations. Their fusion rule can be written as $A  \times A \to 1+\psi$ where $1$ refers to the vacuum and $\psi$ to a fermionic excitation. Only if the fusion outcome is the trivial 1, the pair can either annihilate to the vacuum or tunnel to the next layer. 
In contrast, a fermionic pair will  remain in the plane and the two anyons will further separate due to diffusion. We argue that as long as the system is in the diffusive limit and one is only interested in the real-space dynamics, the only effect of the complex braiding/fusion dynamics is a renormalization of the effective pair-hopping and pair-annihilation rates $\Gamma_\perp$ and $\lambda$ respectively \cite{nahumskinner}.
The corresponding quantum-information dynamics within the internal Hilbert space is also an intriguing problem but beyond the scope of this study.

We analyze the problem numerically using the stochastic particle model of Eq.~\eqref{eq:model}. For an analytical investigation, a coarse grained continuum description of the particle density in the low-density limit is given by the following non-linear diffusion equation for $z\ge 0$
	\begin{align}
		\partial_t\rho = &D_\parallel (\partial_x^2+\partial_y^2)\rho 
		+ D_\perp \partial_z^2\rho^2 - \lambda \rho^2 \label{eq:diffusion_eqn}\\ \nonumber
		&+\eta_\lambda(\vec r,t)+\vec\nabla \vec\xi(\vec r,t) 
	\end{align}
	with a zero-current boundary condition at $z=0$, $(\partial_z \rho^2+\xi_z)|_{z=0}=0$. Comparing to the hopping model, the 
	diffusion constants take the value $D_\|=\Gamma_\| \frac{a^2}{\Delta t} $ with lattice constant $a=1$, time-step length $\Delta t=1$ and $\Gamma_\|=\frac{1}{4}$ in our implementation. Similarly, one obtains $D_\perp=2 \Gamma_\perp \frac{a^2}{\Delta t} $ and $\lambda=2 \Gamma_\lambda \frac{1}{\Delta t}  $, where the factors of $2$ arises since the particle number changes by $2$ (in a layer) during an inter-layer hopping, or an annihilation event.
	The noise arising from the motion of particles is approximately given by $\langle \xi_i(\vec r,t)\xi_j(\vec r',t' )\rangle=\delta_{ij} D_{i} \delta(t-t')\delta(\vec r-\vec r')$
	with $D_{x}=D_{y}=2 D_\| \rho(\vec r,t)$, $D_{z}=4 D_\perp \rho^2(\vec r,t) $ where we fixed the pre-factor of $D_z$ by linearizing the diffusion equations. The noise $\eta_\lambda$ due to pair annihilation is also $\delta$-correlated with the pre-factor $-2 \lambda \rho^2$ \cite{cardy,tauberreview}. The negative sign shows that the noise is not real-valued but has a complex part. 
	
	Besides the continuum model shown in Eq.~\eqref{eq:diffusion_eqn}, we also use a discretized version of this equation using the same lattice as our particle-based simulations, see App.~\ref{app:discretemf}. This allows for a more precise comparison of the models.


	\section{Results}\label{sec:results}
	\begin{figure}
		\centering
		\includegraphics[width=0.9\linewidth]{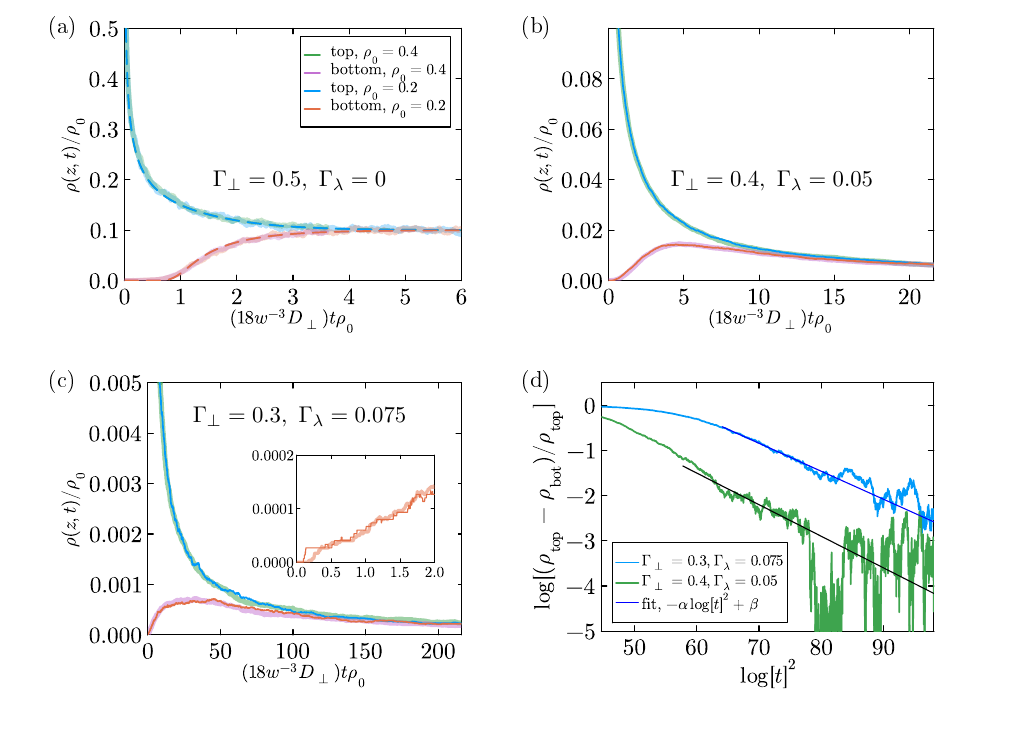}
		\caption{Density evolution in the top and bottom layers of a sample with  $w=10$ layers, where initially all excitations reside in the top layer (using particle-based simulation, Eq.~\eqref{eq:model}). The noisy simulation results is plotted in solid lines while dashed lines in the first panel are obtained by solving the noiseless diffusion equations numerically on a discrete lattice. The collapse of plots for different initial density $\rho_0$ upon rescaling the time and density confirms the scaling predicted by Eq.~\eqref{eq:scaling2}. (a) $\Gamma_\perp=0.5,\Gamma_\lambda=0$   (b) $\Gamma_\perp=0.4, \Gamma_\lambda=0.05$, (c) $\Gamma_\perp=0.3, \Gamma_\lambda=0.15$. All plots are averaged over 6 simulations using a $500\times 500\times 10$ grid. Initial particle densities $\rho_0$ are $0.4$ and $0.2$, corresponding to 100.000 and 50.000 particles, respectively. Panel (d) shows, for two parameter sets with $\rho_0=0.2$, that the difference between particle densities in top and bottom layers decays approximately with $e^{-\alpha \log^2 t}$. Fit parameters: $\alpha=0.062,\beta=3.5$ for the blue points; $\alpha=0.07,\beta=2.7$ for green points.}
		\label{fig:topandbottom}
	\end{figure}
	\subsection{Quench protocol probing topological excitations}\label{sec:quench}
	Our goal is to propose a quench protocol that captures the kinematic constraints and decay processes induced by the robust $2d$ topological order in a $3d$ crystal.
	For this purpose, we inject a finite density $\rho_0=\rho(t=0)$ on the surface of a layered sample, e.g, by using an intense laser pulse - or a THz pulse exciting resonantly pairs of anyons, see App.~\ref{app:nontopo} for a discussion on surface sensitive pumps and the role of non-topological excitations. For simplicity, we assume that the initial density is confined to the top layer at $z=0$, but all qualitative results will be the same when a few top layers are excited. Relaxation to equilibrium at a temperature $T_0\ll \Delta$ takes place through the three different channels given by Eq.~\eqref{eq:model}. 
	
	How do we experimentally probe the above described dynamics driven by pair-diffusion and pair-annihilation? 
	Here, one has to show experimentally that both processes are quadratic in the density of excitations. As the excitation density $\rho_0$ is directly determined by the intensity of the excitation pulse, the initial density can be easily controlled.
	
	We can use a simple scaling argument to obtain how $\rho_0$ enters the non-linear diffusion equation.
	If we use the following rescaling
	\begin{align}
		&   (x,y,z) \to (x b^{-1},yb^{-1},z),\quad t \to tb^{-2}, \quad  \rho\to \rho b^{2} \nonumber \\
		& (\xi_x,\xi_y,\xi_z) \to (\xi_x b^{3},\xi_y b^{3},\xi_z b^{4}), \quad \eta \to \eta b^4.\label{eq:scaling2}
	\end{align} 
	we find that in Eq.~\eqref{eq:diffusion_eqn}  {\em all} $b$-dependencies drop from the equation {\em and} the noise correlators (changing only the short-distance cutoffs). Setting $b^2=\rho_0^{-1}$ and assuming that the initial excitations are uniform in the $xy$ plane, this version of scaling suggests that
	\begin{align}
		\frac{ \rho(z,t)}{\rho_0}=\tilde \rho(z, t \rho_0), \label{eq:scalingRho}
	\end{align}
	where the scaling function $\tilde \rho$ is {\em independent} of $\rho_0$. Note that this is in general {\em not} an exact statement due to extra  corrections arising from the cutoff-dependence of observables giving rise to extra logarithmic corrections, see Sec.~\ref{sec:noise}. In our numerics, see Fig.~\ref{fig:topandbottom}, we do, however, find that  our stochastic particle-based simulations (unexpectedly) obey the simple scaling relation with high precision. In the regimes explored by us we have not been able to identify corrections to the scaling prediction Eq.~\eqref{eq:scalingRho}.

	To explore this physics experimentally, one has to compare the time evolution of observables for different values of $\rho_0$ (controlled by the  intensity of the exciting pulse and, possibly, the initial temperature).
	
	Consider, for example, the density of excitations on the top- and bottom layer after an exciting laser or THz pulse, see Fig.~\ref{fig:topandbottom}. 
	All time scales, e.g., the time scale on which density in the bottom layer rises or that in the top layer drops, are according to Eq.~\eqref{eq:scalingRho} inversely proportional to $\rho_0$ and therefore to the intensity of the exciting pulse, $I_P$. 
	\begin{align}
		\tau \propto \frac{1}{\rho_0}\propto \frac{1}{I_P}\label{eq:scalingTime}
	\end{align}
	This peculiar intensity dependence of all time scales is the smoking-gun signature of the fact that our excitations are topological.
	This scaling property can thus be used to prove experimentally that the physics of a given observable is dominated by pair annihilation and pair diffusion of topological excitations, providing a relatively direct experimental proof of the topological nature of excitations.
	
	In the following, we investigate in detail the dynamics of the cloud of excitation and the validity of the mean-field picture used in the argument above.
	
	

	\subsection{Pair diffusion}\label{sec:pair}
	We first consider the situation where the total energy of the topological excitations is conserved, assuming that, e.g., the coupling to phonons can be neglected. We furthermore assume that the bandwidth $W$ of the excitations is small compared to their gap $\Delta$, $W \ll \Delta$.  Thus the energy of $n$ excitations is approximately $n \Delta$. By energy conservation the number of particles is therefore approximately conserved and there is no pair annihilation, $\lambda =0$.
	
	We assume as discussed in Sec.~\ref{sec:quench} the density is approximately uniform within a layer, $\rho(\vec r, t)=\bar{\rho}(z,t)$. Thus, the diffusion equation \eqref{eq:diffusion_eqn} in the absence of noise and  for $\lambda=0$  becomes
	\begin{align}
		\partial_t\bar{\rho} = D_\perp \partial_z^2 \bar{\rho}^2.
	\end{align}
	We consider a semi-infinite system, $w=\infty$.
	Non-linear diffusion equations of similar kind have been  widely studied \cite{porousmediareview}.
	A standard solution strategy  is based on a scaling ansatz of the form \cite{vazquez2014barenblatt}
	\begin{align}
		\bar{\rho}(z,t) = \frac{F(z/t^\alpha)}{t^\alpha} \label{scaling_ansatz}.
	\end{align}
	
	Plugging this ansatz into the non-linear diffusion and solving for $\alpha$ gives $\alpha=1/3$ and
	\begin{align}\label{eq:scalingSol1}
		F(u) = \frac{u_0^2 - u^2}{12 D_\perp} \theta(u_0^2-u^2),\quad u_0 = (18 D_\perp \rho_0)^{1/3}
	\end{align}
	where $u_0$ is obtained from the initial density $\rho_0$ (number of particles per area in the first layer).
	
\begin{figure}[t]
		\centering
		\includegraphics[width=0.9\textwidth,trim=1cm 0cm 0cm 0cm,clip]{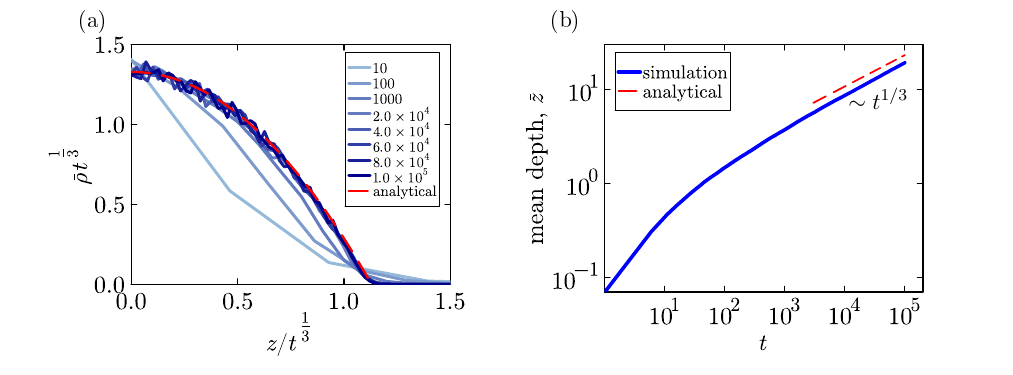}
		\caption{(a) Layer density of excitations along the $z$ direction plotted in rescaled coordinates. The blue curves show snapshots at various times. A scaling collapse happens for long time scales, consistent with the  the analytical prediction of the noiseless diffusion model (red-dashed curve). (b) The average depth $\bar z$ traversed by the excitations into the bulk as a function of time (in log-log scale). The expected scaling of $\bar{z}\sim t^{1/3}$ is shown by the dashed line. Simulation parameters: $L=300,\rho_0=0.1,\Gamma_\perp=0.4,\Gamma_\|=1$. Fit (in (a)):$\rho_0=0.1,D_\perp =0.8$}
		\label{fig:densityprofile}
	\end{figure}

	Thus the excitations penetrate into the bulk of the sytem sub-diffusively.
	The front of the excitations is located at \begin{align}
		Z(t)=u_0 t^{1/3} \label{eq:diffExp}
	\end{align} while the center of mass $\bar z  =\int_{0}^{\infty}dz~ z~\frac{\bar{\rho}}{\rho_0}$ of the excitation cloud is located at  $\frac{3}{8} u_0 t^{1/3}$.
	Fig.~\ref{fig:densityprofile} shows that in the long-time limit the density profile of our particle-based simulations takes the universal parabolic form described by Eq.~\eqref{eq:scalingSol1} which does not contain any fitting parameters. As we will discuss below, the accuracy of the fit to the noiseless mean-field model arises because noise is an irrelevant perturbation in this case.
	
	The excitation density in the top-layer also decays with a sub-diffusive power law
	\begin{align}
		\rho(z=0) \sim \left(\frac{\rho_0^2}{D_\perp t}\right)^{1/3}.
	\end{align}
	\subsection{Interplay of pair-diffusion and pair-annihilation}\label{sec:interplay}
	Next, we consider the case when dissipative processes are present and the particles are allowed to annihilate each other in a process where the excess energy is transferred, e.g., to phonon excitations. Within our microscopic model, such processes occur with a rate $\Gamma_\lambda=\lambda/2$ when two excitations are nearest-neighbors within a layer. The (coarse-grained) non-linear equation in the absence of noise is thus given by 
	\begin{align}
		\partial_t \bar \rho=D_\perp \partial_z^2 \bar \rho^2 -\lambda   \bar \rho^2 .\label{eq:noiseless}
	\end{align}
	
	An asymptotic solution of Eq.~\eqref{eq:noiseless} can be obtained as a closed form (see Ref.~\cite{handbookPDE,Ben-Jacob01062000,Ngamsaadpre} and references therein for related literature.)
	
		\begin{align}	\label{eq:decay_profile}
			\bar{\rho}(z,t) = &\frac{1}{\lambda (t+t_0)}\left[ 1-\frac{e^{\frac{z-z_0}{2}\sqrt{\lambda/D_\perp}}}{2\sqrt{ D_\perp a^{-5}(t+t_0)}}\right] \theta\left(Z(t)-z\right),\\ \nonumber 
			Z(t)=&z_0+\sqrt{\frac{D_\perp}{\lambda}}\log[4 D_\perp a^{-5}(t+t_0)]
		\end{align}
	where $z_0$ and $t_0$ depend on the shape (height and width) of the initial density profile. The factor $a^{-5}$, where $a$ is the lattice spacing, has been introduced purely for dimensionality reasons but has no consequence to the solution since a change of $a$ can be absorbed into $z_0$. Although the above solution does {\em not} obey the
	boundary condition $\partial_z \rho^2=0$ at $z=0$, we have confirmed via numerically solving Eq.~\eqref{eq:noiseless} that it nevertheless accurately describes the solution at long times.
	\begin{figure}[t]
		\centering
		\includegraphics[width=0.7\textwidth,trim=0cm 0cm 0cm 0cm,clip]{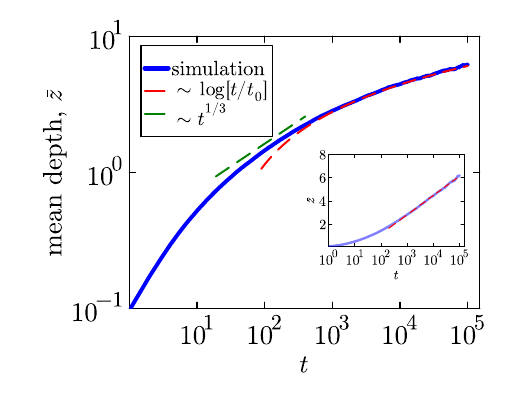}
		\caption{The average depth $\bar z$ traversed by the excitations into the bulk when pair-annihilation processes are present, $\lambda\ne0$. We find the exact simulation results to be consistent with the predictions of Eq.~\eqref{eq:crossover}. At early times, the simulation data is roughly consistent with the sub-diffusive $t^{1/3}$ power-law (green dashed line), which crosses over to a logarithmic scaling at longer times (red dashed line, $t_0=20$). Inset: Data is plotted in log-linear scale to show $\bar{z}\sim \log t$ scaling at long times. Simulation parameters: $L=500,\rho_0=0.2,\Gamma_\|=1,\Gamma_\perp=0.3,\Gamma_\lambda=0.1$.}
		\label{fig:com_anni}
	\end{figure}
	Perhaps counter-intuitively, Eq.~\eqref{eq:decay_profile} shows that even a weak pair-annihilation suppresses the propagation of the density cloud. Instead of $\bar z\sim t^{1/3}$ we now obtain  the mean depth $$\bar z\sim \log t.$$ At the same time, the surface density drops asymptotically as $\rho(0)\sim1/t$,  much faster than $1/t^{1/3}$ obtained for $\lambda=0$. Thus, the total number of particles $n$ decays as $\log t/t$ at long times, within the mean field (noiseless) approximation. The extra $\log t$ factor arises as pair annihilation is suppressed due to the (logarithmic) expansion of the cloud in $z$ direction resulting in reduced particle densities. 
	
	For small $\lambda$, we expect a crossover from a regime dominated by pair diffusion at short time scales to the annihilation-dominated regime discussed above. We estimate the crossover time scale using the scaling solution Eq.~\eqref{eq:scalingSol1} and the condition that the two terms on the right-hand side of Eq.~\eqref{eq:noiseless} are of comparable size. Thus, for the width of the cloud (up to logs in the crossover scale), we obtain 
	\begin{align}
		Z(t) \approx \left\{\begin{array}{ll}
			(D_\perp \rho_0 t)^{1/3}   &  \text{for } t \ll \frac{\sqrt{D_\perp}}{\lambda^{3/2} \rho_0}\\
			\sqrt{\frac{D_\perp}{\lambda}}\log{t} &  \text{for } t \gg \frac{\sqrt{D_\perp}}{\lambda^{3/2} \rho_0}
		\end{array}\right.
		\label{eq:crossover}
	\end{align}
	This crossover is not captured by the analytical solution which is only valid in the long-time limit.
	
	In Fig.~\ref{fig:com_anni}, we show that this analytical prediction is consistent with our numerical results.

	\subsection{Corrections from noise}\label{sec:noise} We will now examine how various sources of noise affect the results of the noiseless diffusion equation discussed above.
	
	We first consider the case without annihilation,  $\lambda=0$. To estimate the effect of noise, we perform a scaling analysis of Eq.~\eqref{eq:diffusion_eqn} based on its asymptotic solution in the absence of noise. First, we  rescale the variables using the following transformations
	\begin{align}
		&(x,y,z) \to (x b^{-1},yb^{-1},zb^{-2/3}),\quad t \to tb^{-2} \nonumber \\
		&\rho\to \rho b^{2/3},  \quad   (\xi_x,\xi_y,\xi_z) \to (\xi_x b^{8/3},\xi_y b^{8/3},\xi_z b^{3}).\label{eq:scaling1}
	\end{align} 
	Here, the scaling of space, time and $\rho$ follows from Eq.~\eqref{scaling_ansatz}. The scaling exponents for the noise are chosen to make the noise correlator $b$ independent, using that  $\langle \xi \xi \rangle$  scales with $b^{-1-1-2/3-2-2/3}=b^{-16/3}$ for the in-plane noise and $b^{-6}$ for the out-of-plane component.
	
	Expressing all the terms in Eq.~\eqref{eq:diffusion_eqn} (for $\lambda=0$) in the rescaled coordinates and fields, and multiplying with $b^{8/3}$, we find that the pre-factors of both
	$\nabla_x \xi_x+\nabla_y \xi_y$ 
	and 
	$\nabla_z \xi_z$ are suppressed by $1/b$.  This shows that noise is irrelevant at long times for $\lambda=0$. This is also confirmed  by our exact particle-based simulations, the results of which are well-described by the analytical solution of the noiseless model, see Fig.~\ref{fig:densityprofile}.
	

	For $\lambda\ne 0$, we instead use the scaling analysis discussed in Eq.~\eqref{eq:scaling2}. While we used it in Sec.~\ref{sec:quench} to obtain the exact dependence of the solutions on $\rho_0$, here we use the exact scale invariance of Eq.~\eqref{eq:diffusion_eqn} to argue that noise is a {\em marginal} perturbation, implying that it cannot simply be neglected. 
	Indeed, in our simulations with $\Gamma_\lambda\ne 0$, we find deviations from the mean-field predictions in the long-time limit.

	\begin{figure}[t]
		\centering
		\includegraphics[width=0.9\textwidth,trim=0cm 0cm 0cm 0cm,clip]{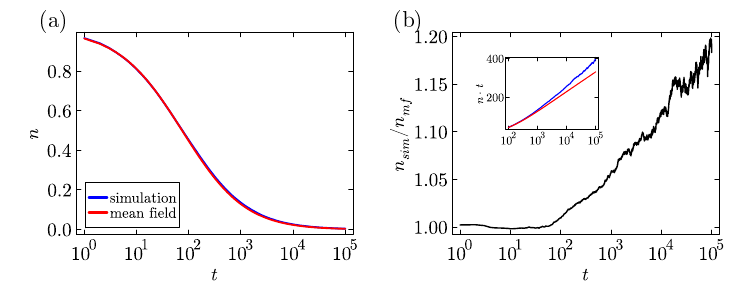}
	\caption{(a). Total density $n$ (particles per area) as a function of time (shown in log-linear scale), for $\Gamma_\perp=0.3, \Gamma_\lambda = 0.1$, starting from an initial density $\rho_0=0.2$. The red curve is obtained by numerically solving the noiseless diffusion equation (mean-field) on a grid of size $100$. (b). The ratio between the solution $n_{\text{mf}}(t)$ obtained from the noiseless model and exact simulation $n_{\text{sim}}(t)$ are plotted, which shows an approximately $n_\text{sim}/n_\text{mf}\sim \log t$ behavior for $t\gtrsim 10^2$. Inset: Plot of $t\,n(t)$, showing that $n_\text{sim}\sim (\log t)^2/t$ while  $n_\text{mf}\sim (\log t)/t$. Simulation parameters are identical to that of Fig.~\ref{fig:com_anni}.}
	\label{fig:noise}
\end{figure}

In the absence of pair diffusion, $D_\perp=0$, the problem of pair-annihilation has been widely studied \cite{cardy}. In this $2d$ case, the marginal nonlinear coupling $\lambda$ turns out to be marginally irrelevant, effectively decaying with $1/ \log t$ \cite{cardy,tauberreview}. As within mean-field $n(t)\sim 1/(\lambda t)$,
the particle density therefore decays with $n_{2d}(t)\sim \log{t}/(D_\parallel t)$.
The logarithmic enhancement arises from a logarithmic increase of the probability of a diffusive particle to come back to its origin, which leads to logarithmic suppression of probability to diffuse to the location of a different particle. For $d<2$, one instead obtains from the same mechanism power-law corrections.

Previously, we found that within the mean-field theory, the cloud expands very slowly with $Z(t)\sim \sqrt{D_\perp/\lambda} \log t$.
Combining this slow logarithmic expansion in the $z$ direction with the $2d$ result, it suggests that -- up to multiplicative factors of order $\log(\log t)$ -- the total density will decay as 
\begin{align}
	n(t) =\int d\vec r \rho(\vec r, t)\sim \frac{(\log t)^2}{t}.
	\label{eq:decay_law}
\end{align}
This result is consistent with our numerical simulations, as shown in Fig.~\ref{fig:noise}. Deviation from mean-field are best visualized, see Fig.~\ref{fig:noise}b,  by plotting the ratio of the total particle number obtained from exact simulations and mean-field (noiseless) equations or by plotting the product $n(t) t$ (inset). Note, however, that for the shown parameters and time scales, the noise-induced logarithmic corrections are only on the level of $20\,\%$.


\subsection{Finite slab geometry and experimental signatures}
\label{sec:finiteslab}
Above, we discussed the time-evolution in a half-infinite system. For an experimental implementation, considering a finite slab has substantial advantages.

We consider the following setting:
After an excitation on the top surface (e.g., by a laser or THz pulse), one tracks the density of excitations on both the top and bottom surface of a slab of width $w$ as function of time and, importantly, the intensity of the exciting pulse. As discussed in Sec.~\ref{sec:quench}, the density dependence of all time scales, Eq.~\eqref{eq:scalingTime}, is the smoking gun signature of topological excitations.

For a slab of width $w$, two different regimes arise. For $w\ll \sqrt{D_\perp/\lambda}$, the physics is dominated by pair diffusion, see Eq.~\eqref{eq:scalingSol1}, while annihilation governs the opposite regime.
According to our previous analysis, the excitation reach the lower layer after time 
\begin{align}
	t_b \sim \left\{
	\begin{array}{ll}
		\frac{w^3}{18 D_\perp \rho_0}, & \quad w\ll \sqrt{D_\perp/\lambda}\\[2mm]
		e^{w \sqrt{\lambda/D_\perp}},   & \quad w\gg \sqrt{D_\perp/\lambda}
	\end{array}
	\right. .
\end{align} 
Note that the scaling relation $t_b \propto 1/\rho_0$ is hidden in the pre-factor of the exponential in the large $w$ regime, which depends on the details of the initial density profile.
For $t\to \infty$ the density becomes uniform in the $z$ direction and thus the density of top and bottom layer approach each other. 
To compute analytically how the density at top and bottom layer approach each other, we Taylor expand around the uniform solution of the differential equation, see Appendix~\ref{app:slab}. Calculating the leading correction, we obtain
\begin{align}
	\frac{ n_\text{top}-n_\text{bottom}}{n_\text{top}}\sim e^{-\alpha(\log t)^2}.
\end{align}
For 
$w\gg \sqrt{D_\perp/\lambda} $ the exponent $\alpha\approx
\frac{\lambda w^2+2\pi^2D_\perp}{16\pi w^3 D_\|}$ 
becomes small. Thus, the densities at top and bottom surfaces approach each other very slowly in this regime, as confirmed by our numerical results, see Fig.~\ref{fig:topandbottom}.

{ \color{black} \subsection{Ultraclean limit}
So far, we considered the scenario where the in-plane dynamics is diffusive with a density-independent diffusion constant $D_\|$. Here, scattering from impurities (disorder) or phonons may typically determine $D_\|$. From a theoretical perspective, it is also interesting to explore what happens in the absence of disorder and at temperatures so low that phonons can be neglected. In this case, the only source of scattering is collision between the topological excitations with each other.  Generically, we expect that this inter-particle scattering rate is much larger than both the particle-particle annihilation rate (which is only possible due to phonon-assisted processes) and the pair-hopping rate (proportional to the interlayer coupling). This justifies the use of an emergent diffusion equation for the in-plane motion, as the excitations scatter many times before hopping to neighboring layer. We also assume that the initial laser pulse excites particles with a wide distribution of momenta so that Umklapp scattering processes can occur. As other particles are the only source of scattering, the diffusion constant is inversely proportional to their density, $D_\|=\frac{\tilde D_|} {\rho}$ \cite{schneider2012fermionic,rapp}, and instead of Eq.~\eqref{eq:diffusion_eqn} we obtain
	\begin{align}
		\partial_t\rho &=   \tilde D_\parallel \left(\partial_x \left(\frac{1}{\rho} \partial_x \rho\right)+\partial_y \left( \frac{1}{\rho} \partial_y \rho \right) \right)
		+ D_\perp \partial_z^2\rho^2 - \lambda \rho^2 \label{eq:diffusion_eqn_ball}\\ \nonumber
		&+\eta_\lambda(\vec r,t)+\vec\nabla \vec\xi(\vec r,t) 
	\end{align}
where the noise correlators are also modified ($D_x=D_y=2 \tilde D_\|$). Such a singular diffusion equation has profound impact on the in-plane dynamics when the initial density is non-uniform within the $xy$ plane; see Ref.~\cite{schneider2012fermionic} for a combined theoretical and experimental study in an ultracold-atom system.
In this case, the particle cloud expands with a diffusive core surrounded by ballistically moving front.

That said, in our study, we consider only the case where the system is uniformly excited in the $xy$ plane. Therefore, in the noiseless limit, the in-plane diffusion term drops completely from our equation. Nevertheless, one needs to examine the effect of noise using the scaling analysis of Sec.~\ref{sec:noise} for Eq.~\eqref{eq:diffusion_eqn_ball}. For $\lambda=0$ we thus obtain, instead of Eq.~\eqref{eq:scaling1},
	\begin{align}
	&(x,y,z) \to (x b^{-4/3},yb^{-4/3},zb^{-2/3}),\quad t \to tb^{-2} \nonumber \\
	&\rho\to \rho b^{2/3},  \quad   (\xi_x,\xi_y,\xi_z) \to (\xi_x b^{8/3},\xi_y b^{8/3},\xi_z b^{10/3}),\label{eq:scalingballistic1}
\end{align} 

while for $\lambda>0$ we find, instead of Eq.~\eqref{eq:scaling2},
	\begin{align}
	&   (x,y,z) \to (x b^{-2},yb^{-2},z),\quad t \to tb^{-2}, \quad  \rho\to \rho b^{2} \nonumber \\
	& (\xi_x,\xi_y,\xi_z) \to (\xi_x b^{3},\xi_y b^{3},\xi_z b^{5}), \quad \eta \to \eta b^5.\label{eq:scalingballistic2}
\end{align}
In both cases it turns out that the noise is {\em irrelevant}. Physically, the increased in-plane diffusion makes the system more homogeneous, suppressing the effect of noise. Therefore, we can directly use the results from Sec.~\ref{sec:pair} and Sec.~\ref{sec:interplay} to describe the ultraclean low-T limit, as long as the surface is uniformly excited.
}

\section{Discussions and Conclusion}

Spin liquids and other phases of matter with intrinsic topological order are  -- aside from the notable exception of quantum Hall phases --  notoriously difficult to detect. The core challenge is that topological order, by its very nature, cannot be identified by a  local order parameter. As a result, one must instead rely on indirect signatures such as thermodynamic responses, heat transport, or the observation of a continuum of excitations in spectroscopic measurements. In particular, the often unavoidable presence of disorder makes the unambiguous identification of spin-liquid phases especially challenging.

Here, we explore an alternative route to detecting topological order in layered two-dimensional materials. In these systems, topology enforces that excitations carrying gauge charge are confined to move within individual layers. In the $\mathbb Z_2$
model considered here, inter-layer motion is possible only for pairs of such excitations. We argue that suitably designed pump–probe experiments can directly probe the effective dimensionality of the excitations, thereby offering a potentially robust signature of the underlying topological order.

Here, the most striking signature is that, for uniform excitations of the system, all relevant time scales are inversely proportional to the density of excitations. Our analysis has also shown that pair annihilation is very effective in slowing down the propagation of topological excitations along the direction perpendicular to the layers. In a semi-infinite system this leads to a slow logarithmic expansion in the $z$ direction. In a finite slab, in contrast, the density of top and bottom layer approach each other very slowly, following a stretched exponential in logarithmic time, $e^{-\alpha (\log t)^2}$.


A key prerequisite of our detection protocol is the availability of a probe sensitive to the density $\rho$ of the topological excitations. The optimal choice of probes will depend on the microscopics of the system, but in general, one expects that many observables—such as the dielectric function \cite{opticalherbertsmithite,banerjee2023electromagnetic,haooptical}, Raman intensities \cite{raman_u1,joyRaman}, and others—depend approximately linearly on $\rho$.

Similar pump-probe schemes can also be used to explore other phases of matter where 
single excitations can move only in a $d'$ dimensional subspace of a $d$-dimensional system, $d'<d$. This includes fracton-like phases \cite{fractonReview2019} in quantum liquids and -- in purely classical systems -- the dynamics of dislocations in crystals or charge density waves \cite{topdefectsprx25,mitrano2019ultrafast}. 
An interesting question is to investigate the quantum dynamics and entanglement growth of non-Abelian anyons with internal degrees of freedom. Incorporating their braiding and quantum mechanical interactions into our framework is left for future work.

In our model and similar systems, the effective dimensionality of single-particle excitations is protected by topology. Interestingly, this changes in the presence of a finite density of screw dislocation as pointed out in Ref.~\cite{floatingtop,screwfractons}. By encircling a screw dislocation a single topological excitation can move from one layer to the next. Thus, screw dislocations effectively act as a set of spiral staircases connecting the different layers, allowing for single-particle diffusion in the $z$ direction.

In conclusion, we propose using pump–probe experiments to investigate the effective dimensionality of  topological excitations, applicable to a variety of systems. Although we have focused on a scheme in which the entire top layer is illuminated uniformly, the analysis can be readily extended to cases where only a finite spot on the surface is excited. In such configurations, one can simultaneously track the propagation of excitations both within the plane and perpendicular to it.
\section*{Acknowledgements}
	We thank Ajesh Kumar, Sebastian Diehl and Urban Seifert for useful discussions.
\paragraph{Funding information}
This work was supported by the Deutsche Forschungsgemeinschaft (DFG) through CRC1238 (Project No.\ 277146847, projects C02 and C04).
\paragraph{Data availability}
The numerical data presented in this work is available at zenodo.org\cite{joy_2025_17880707}
\begin{appendix}
\numberwithin{equation}{section}
\section{Mean-field solution on a lattice}\label{app:discretemf}
To solve the mean-field noiseless model, we implement descretize Eq.~\eqref{eq:noiseless} on a $1d$ lattice along the $z$ direction with length $w$. For each layer, we then obtain
\begin{align}
	\partial_t\rho_l=&D_\perp \left( \rho_{l+1}^2-2\rho_l^2+\rho_{l-1}^2\right)-\lambda \rho_l^2,\quad l=0,1,...w-1 \\ \nonumber
	\partial_t\rho_0=&D_\perp \left( \rho_{1}^2-\rho_0^2\right)-\lambda \rho_0^2 \\ \nonumber
	\partial_t\rho_w=&D_\perp \left( \rho_{w-1}^2-\rho_w^2\right)-\lambda \rho_w^2,
\end{align}
where $\rho_l$ is the density on layer $l$.
The last two lines above implement the boundary conditions at $z=0$ and $z=w$. We solve this system of equations using Runge-Kutta method with the initial condition $\rho_l(t=0)=\rho_0\delta_{l,0}$.
\section{Perturbation analysis for finite slab}\label{app:slab}
In a finite slab of a layered sample with thickness $w$, the rate with which densities of the top and bottom layers approach each other, after an initial excitation on the top layer, can be estimated using a perturbation analysis around the uniform solution to the diffusion equation Eq.~\eqref{eq:noiseless}.

In the limit $t\to\infty$, we expect the density to be independent of $z$, denoted by $\rho_0(t)$. For example, within the noiseless model Eq.\eqref{eq:noiseless}, $\rho_0(t)=(\lambda t)^{-1}$. We consider a small perturbation of the form $\bar{\rho}(z,t)\approx \tilde{\rho}_0(t)+\delta \rho(z,t)$ where $\delta\rho/\tilde{\rho}_0\ll1$, and obtain at first order in $\delta\rho$
\begin{align}\label{eq:lin}
	\partial_t \delta\rho=2\bar\rho_0(t)\left(D_\perp \partial^2_z\delta\rho-\lambda \delta\rho\right)
\end{align}
Using the Fourier series expansion $$\delta\rho(z,t)=\sum_{n=1}^\infty \delta\rho_{k_n}(t)\cos(k_nz),$$ where the wave-vectors $k_n=\pi n/w$ with $n\in Z$, we can obtain the solution for the components $\delta\rho_n$. Note that only the cosine terms appear in the expansion so that our boundary conditions of no currents at $z=0$ and $z=w$ are imposed. Therefore, we obtain
\begin{align}
	\delta\rho_{n}(t) = \delta\rho_n(0) \exp\left[{-2\left[\frac{D_\perp n^2\pi^2}{w^2}+\lambda\right]\int_0^t\bar\rho_0(\tau)}\right], \label{eq:decay_n}
\end{align}
where $\delta\rho_n(0)$ are constants determined by the shape of the profile at time scales where both the top and bottom layer densities start to decay, see Fig.~\ref{fig:topandbottom}b.

We thus obtain the difference between the top and bottom densities 
$$\Delta(t)=\left(\bar{\rho}(0,t)-\bar{\rho}(w,t)\right)/\bar{\rho}(0,t)$$ at leading order in $\delta\rho$
\begin{align}
	\Delta &\approx \sum_{n=1}^{\infty}\frac{\left(\delta\rho_n-(-1)^n\delta\rho_n\right)}{\rho_0(t)}\\ \nonumber
	&=\sum_{n=1,3,5,...}^{\infty}\frac{2\delta\rho_{n}(0)}{\rho_0(t)}\exp\left[{-2\left[\frac{D_\perp n^2\pi^2}{w^2}+\lambda\right]\int_0^t\rho_0(\tau)}\right]
\end{align}
Since larger $n$ modes are exponentially suppressed and $\delta\rho(z,0)$ is assumed to be a smooth function, we can obtain the leading behavior by truncating the sum to $n=1$,
\begin{align}
	\Delta(t) \propto \frac{e^{-2\left(\frac{D_\perp \pi^2}{w^2}+\lambda\right) \int_0^t \rho_0(t)}}{\rho_0(t)},
\end{align}
For the noiseless (mean-field) case, $\rho_0(t)=(\lambda t)^{-1}$ and we obtain 
\begin{align}
	\Delta_{\text{mf}} \sim \lambda t^{-\alpha}, \quad \alpha=1+\frac{2D_\perp \pi^2}{\lambda w^2}\label{eq:aniMF}
\end{align}

For the noisy case, one can approximately obtain the leading behavior by replacing $\rho_0(t)$ with the steady state solution for a $2d$ diffusion-annihilation model. This is given by  the well-known \cite{cardy} formula $\rho_0(t)=\frac{\log(D_\|t/a^2)}{8\pi w D_\| t}$ in the long-time limit (note the extra factor of $1/w$ compared to \cite{cardy} arising as $\rho$ is a $3d$ rather than the $2d$ density). This solution is independent of $\lambda$ as annihilation processes are controlled by the in-plane diffusion as the particles become more and more dilute at long times. Thus we obtain the leading behavior for the density difference
\begin{align}\label{eq:aniDecay}
	\Delta(t)\sim \frac{ t}{\log t} e^{-\alpha(\log t)^2}.
\end{align}
From our derivation, we obtain 
$\alpha= \frac{\lambda w^2+2\pi^2D_\perp}{16\pi w^3 D_\|}$ but this formula does not take into account possible renormalizations of $\lambda$ and $D_\perp$ in Eq.~\eqref{eq:decay_n} arising from non-linear interactions of $\delta \rho_1$ with other Fourier modes with $n\neq 0$. 
For the finite-width system only the $n=0$ mode obtains logarithmic corrections and correspondingly these renormalizations  remain finite. Thus, in principle, $\alpha$ in Eq.~\eqref{eq:aniDecay} should be viewed as a fitting parameter. Numerically, we find that $\alpha$ becomes small in the pair-annihilation dominated regime, $\lambda w^2 \gg D_\perp$, as predicted by the mean-field result, Eq.~\eqref{eq:aniMF}. Note, however, that most of our numerics is not in the asymptotic regime dominated by logarithmic corrections as can be seen from Fig.~\ref{fig:noise} where logarithmic corrections remain smaller than $1$.

\begin{figure} 
	\centering
	\includegraphics[width=0.8\textwidth,trim=0cm 0cm 0cm 0cm,clip]{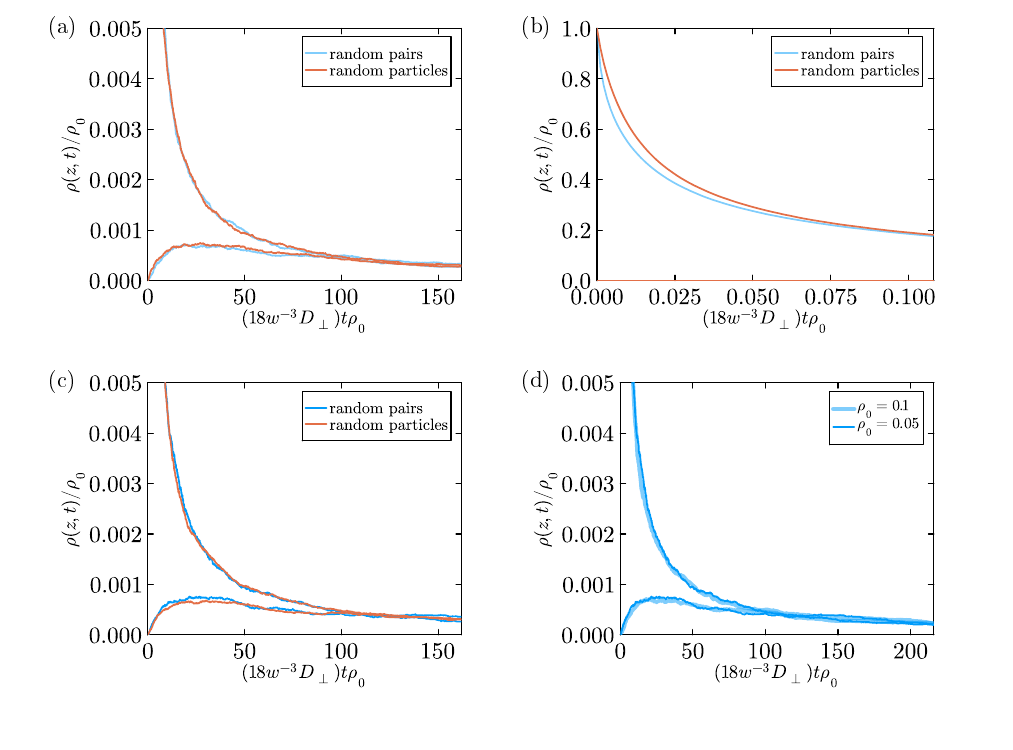}
	\caption{\label{fig:doubles} (a) Density evolution of the top and bottom layer for an initial state with randomly placed nearest-neighbor pairs (blue) and randomly placed particles (red), with the same initial density $\rho_0=0.1$. While the paired initial condition leads to a slightly faster decay of the top layer density at early times (see (b)), they converge rather quickly onto each other. In (c), we confirm the scaling law Eq.~\eqref{eq:scalingTime} for random-pairs initial conditions. The simulations are performed on a lattice grid of size $1000\times 1000\times 10$, and model parameters $\Gamma_\lambda=0.3$ and $\Gamma_\lambda=0.15$. Initial density of $0.1$ and $0.05$ correspond to $50,000$ and $100,000$ particles respectively.}
\end{figure} 
\section{Effect of initial correlations}\label{app:initial}

In Fig.~\ref{fig:doubles} we explore the effects of pair correlations in the initial state. We compare two initial conditions: 1.  randomly placed particles in the first layer and 2. randomly placed nearest-neighbor pairs of particles. In both cases, we do not allow configurations where two particles occupy the same site.
The second initial condition takes into account that our topological excitations can be locally created only in pairs.  At very short times, see Fig.~\ref{fig:doubles}b, the pair annihilation rate is slightly enhanced compared to  the second initial condition. For longer times, however, the pair correlations of  the initial state have no visible effects.

\section{Role of non-topological excitations and experimental considerations}\label{app:nontopo}
The analysis presented in the main part of the paper assumes that the only relevant excitations in the system are topological. In a real material , however, the presence of non-topological excitations such as phonons and photons is unavoidable. 
In this appendix we give a brief, qualitative discussion on how they affect our results and discuss under what conditions it is possible to create excitations strictly close to the surface of a material.

When using a THz or laser pulse, its penetration depth into the sample is usually determined by the absorption rate. Therefore, one has to choose a frequency range where absorption (either directly by pairs of topological excitations, or by other degrees of freedom) is sufficiently large. Alternatively, one may add extra coating layers made from a material with strong absorption or with a large dielectric constant \cite{fish2009total}. In the latter case, illumination from the high-$\epsilon$ side at shallow incidence can generate total internal reflection and an evanescent field near the interface. 

The frequency of the exciting pulse will determine whether it will create primarily the topological excitations  (for frequencies close to twice the gap $\Delta$), or other electronic or phonon excitations which then decay into the low-energy topological excitations (and extra low-energy phonons) only in later stages. In the former scenario, the analysis presented in our paper directly applies (possibly with small modifications from longer-ranged hopping processes discussed below). In the later case, where non-topological excitations dominate initially, the our analysis  only applies on time-scales long compared to the time-scale required to convert high-energy excitations down to low-energy topological excitations and, possibly, low-energy phonons. If these processes are slow enough, they may mask the physics discussed by us.

Another interesting effect, not taken into account in our analysis is the following: a pair of topological excitations may annihilate in one layer, creating a phonon which is reabsorbed in a different layer, where it creates another pair of topological excitations. This results in a phonon-mediated long-ranged pair-hopping process. While such processes will have a very small prefactor (quadratic in the effective phonon coupling), they can still dominate long-distance transport as we have shown that the usual diffusive process are highly ineffective in the presence of pair annihilation. Importantly, phonon-mediated pair hopping will obey the same scaling relations used in Eq.~\eqref{eq:scaling2} of the main text. Therefore the central prediction that all relevant time-scales are inversely proportional to the density of excitations remains valid.

%
\end{appendix}





\bibliography{references.bib}


\end{document}